# FPGA architecture for multi-style asynchronous logic

N. Huot, H. Dubreuil, L. Fesquet, M. Renaudin

TIMA Laboratory, 46, avenue Félix Viallet, 38031 Grenoble – France
Laurent.Fesquet@imag.fr

**Abstract**

*This paper presents a novel FPGA architecture for implementing various styles of asynchronous logic. The main objective is to break the dependency between the FPGA architecture dedicated to asynchronous logic and the logic style. The innovative aspects of the architecture are described. Moreover the structure is well suited to be rebuilt and adapted to fit with further asynchronous logic evolutions thanks to the architecture genericity. A full-adder was implemented in different styles of logic to show the architecture flexibility.*

## 1. Introduction

With the ever-increasing integration level of synchronous design, the industry is now facing problems (heat dissipation, clock tree distribution, noise…) that have led asynchronous logic to gain in popularity these years. Many asynchronous demonstrators have been implemented [1]. The methodologies have also been developed to automatically synthesize asynchronous circuits and dedicated synthesis tools have appeared [2]. Some dedicated FPGAs have also been developed in the last decade to test asynchronous designs. Unfortunately, these FPGAs are closely associated to a style of design. For instance MONTAGE [4] and PGA-STC [5] are based on a synchronous design, GALSA [6] and STACC [7] are globally asynchronous FPGAs but locally synchronous, and PAPA [8] is a fully asynchronous FPGA dedicated to optimize pipeline circuits. The use of synchronous FPGAs is possible but most of the FPGA resources are then unexploited [3]. In this paper, a novel and modular FPGA architecture is presented that is able to implement various asynchronous styles, protocols and data-encodings.

## 2. Principles of Asynchronous Logic

While in synchronous circuits a clock globally controls activity, asynchronous circuit activity is locally controlled using communication channels able to detect the presence of data at their inputs and outputs [9]. This is consistent with the so-called handshaking protocol. Therefore asynchronous modules communicate with each other using requests and acknowledges. One transition on a request signal activates another module connected to it. Therefore, signals must be valid all the time. Hazard is not allowed on signals. Asynchronous circuit synthesis must be thereby more strict, i.e. hazard-free. In fact, different timing assumptions are considered for different types of asynchronous circuits. The most robust style is called Delay Insensitive (DI) because no timing assumption is made. This means that the circuit works correctly whatever the delays are in wires and gates. Having such a circuit is really constraining for the designer and costly in term of area. To reduce the complexity of these circuits, it is possible to introduce an assumption on forks: the forks must be "isochronic" (the delays in the branches of the fork are equal). This style of circuits is named QDI (for Quasi-Delay Insensitive). The "isochronic fork" condition is very weak and many asynchronous circuits have stronger timing assumptions, as micropipeline circuits. The micropipeline circuits only differ from the synchronous circuits by the controllers that replace the clock. Many other asynchronous logic styles exist, but are not presented in this paper. To complete the huge possibilities in asynchronous designs (contrarily to synchronous style), the designer can change the handshake protocol or the data encoding. That means that it is possible to implement asynchronous logic with different protocols or data encoding, like dual-rail (1 of 2 encoding) or multi-rail (1 of N encoding). These choices permit the implementation of a same design varying the electrical properties of the circuit, like speed, power-consumption or electromagnetic emission.

## 3. Architecture

The FPGA architecture has been designed to be the best compromise between the high flexibility needed to be style-independent and the optimal use of FPGA resources. The high flexibility is achieved by choosing an "island style" top view of our chip: the Programmable Logic Blocks (PLBs), which implement the required logical functions, are plunged into a routing network. This network is a grid of interconnection busses, connection boxes, and switch boxes.

The PLB implements the programmable logical functions; it consists of an Interconnection Matrix (IM), two Logic Elements (LE), and a Programmable Delay Element (PDE) as shown in Figure 1. The LEs are programmable combinatorial logic components which host the programmed functions and the PDE gives to the PLB the possibility to implement delayed logic. In addition, the IM maps together PLB inputs, LEs inputs and outputs, and the PDE. The architecture of the PLB is designed to ensure a correct implementation of memory elements typically needed by the asynchronous logic, such as Muller gates [9]. In fact, these memory elements are implemented by mapping looped combinatorial logic using the interconnection matrix integrated into the PLB.

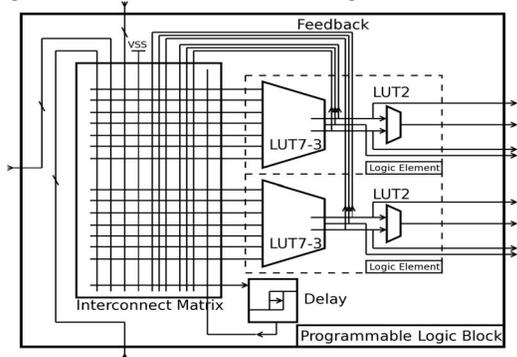

**Figure 1: Internal schematic view of a PLB.**

A LE consists in a "multi-output LUT", a LUT7-3 (7 inputs and 3 outputs), and a LUT2-1 connected together as shown in Figure 2. As presented in Section 2, asynchronous logic uses often 1 of N encoding. This specificity needs to be supported at the hardware level to have the best PLB filling ratio. The adopted solution was to make externally available some internal signals of a LUT; in particular, a LUT7-3 was chosen in the LE. Thus, it becomes easier to implement 1 of N encoding, as auxiliary outputs per LE are available for Multi-Rail signals. Moreover, asynchronous logic also needs to compute the data validity which is used for the protocols; this is supported by adding a LUT2, directly plugged to the multi-output LUT.

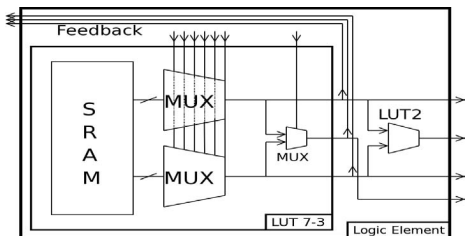

**Figure 2: Internal schematic view of a LE.**

The PDE, located in the PLB (Figure 1) can be used to allow the implementation of asynchronous circuits that need timing assumptions.

## 4. Example

To demonstrate the capabilities of the FPGA architecture, a full adder has been implemented in two different asynchronous logic styles: QDI and micropipeline. In order to simplify the demonstration, the encoding of the QDI adder only is limited to Dual-Rail and the data encoding of the micropipeline adder is bundled data (as in synchronous logic). Moreover, both styles use the same 4-phase protocol.

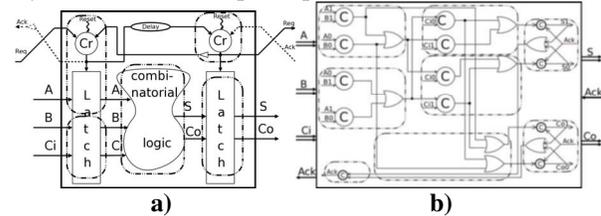

**Figure 3: Full Adder in a) micropipeline and b) QDI.**

Figure 3a and 3b show the micropipeline and the QDI implementation of a 1-bit full adder. The dashed lines around the gates symbolize the mapping in the LEs of the adder. A programmable delay element is used to implement the timing assumption of the micropipeline logic.

## 5. Conclusion and Future Works

A novel FPGA architecture has been presented that is able to target multiple styles of asynchronous logic. The asynchronous logic fits nicely into this dedicated architecture with an overall filling ratio of 51% for the micropipeline circuits and 76% for the QDI circuits. This FPGA circuit will be a tool to evaluate asynchronous designs and to spread this technology to a larger community.